\documentstyle[12pt,epsfig,axodraw]{article}

\newcommand{\ee}{e^+e^-}
\newcommand{\tb}{\tan \beta}
\newcommand{\s}{\smallskip}
\newcommand{\nn}{\noindent}

\newcommand{\beq}{\begin{eqnarray}}
\newcommand{\eeq}{\end{eqnarray}}

\begin{document}
\begin{flushright}
PM--02--42\\
October 2002
\end{flushright}

\vspace*{.5cm} 

\renewcommand{\thefootnote}{*}
\centerline{\bf \large{\sc{SUSY Calculation Tools}}\footnote{Plenary talk given
at SUSY2002 DESY Hamburg, 17--23 June 2002.}}

\vspace*{.5cm}

\renewcommand{\thefootnote}{+}
\centerline{\bf \sc{Abdelhak DJOUADI}\footnote{Present address: TH 
Division, CERN, CH--1211 Geneva 23, Switzerland.}} 

\vspace*{0.3cm}

\centerline{\it Laboratoire de Physique Math\'ematique et Th\'eorique, 
UMR5825--CNRS}
\centerline{\it Universit\'e de Montpellier II, F--34095 Montpellier Cedex 5, 
France.}

\vspace{.5cm}

\begin{abstract}

\nn I discuss the various available tools for the study of the properties of
the new particles predicted in the Minimal Supersymmetric extension of the
Standard Model.  Emphasis will be put on the codes for the determination of the
sparticle and  Higgs boson spectrum. Codes for the calculation of production
cross sections,  decay widths and branching ratios, Dark Matter relic density
and detection rates,  as well as codes for automatic analytical calculations
and Monte-Carlo  event generators for Supersymmetric processes will be briefly 
discussed.

\end{abstract}

\vspace*{.7cm}

\setcounter{footnote}{0}
\renewcommand{\thefootnote}{\arabic{footnote}}
\subsection*{1. Introduction}

It is a well--known fact that in broken supersymmetric theories, it is a
rather  tedious task to deal in an exhaustive way with all the basic parameters
of the  Lagrangian, to derive their relationship with the physical parameters,
the  particle masses and couplings, and to make detailed and complete 
phenomenological analyses  and comparisons with the outcome or expectations 
from experiments. This is mainly due to the fact that, even in the Minimal 
Supersymmetric Standard Model (MSSM) \cite{MSSM,GDR}, with a: 

-- minimal gauge group, the Standard Model ${\rm SU(3)_C \times SU(2)_L 
\times U(1)_Y}$ one, 

-- minimal particle content: three generations of ``chiral" sfermions 
$\tilde{f}^i_{L,R}$ [no right-- \\ \hspace*{.8cm} handed sneutrinos] and two 
doublets of Higgs fields $H_1$ and $H_2$,
 
-- minimal set of couplings imposed by R--parity conservation to enforce 
baryon and \\ \hspace*{.8cm} lepton number conservation in a simple way,

-- minimal set of soft SUSY--breaking parameters: gaugino mass terms $M_i$, 
scalar mass \\ \hspace*{0.8cm} terms $m_{H_i}$ and $m_{\tilde{f}_i}$, a 
bilinear term $B$ and trilinear sfermion couplings $A_i$, 

\nn there are more than hundred new parameters  \cite{parameters} in the
general case of arbitrary complex phases, intergenerational mixing and
non-diagonal sfermion mass and coupling matrices.   Even if one constrains the
model to have a viable  phenomenology, assuming for instance no
intergenerational mixing, no large new source of CP violation, universality of
first and second generation  sfermions [a model that we will call \cite{GDR}
phenomenological or pMSSM], there are still more than 20 free parameters left
to cope with.\newpage

This large number of inputs enters in the evaluation of the masses of ${\cal O}
(30)$ SUSY particles and Higgs bosons as well as their complicated couplings, 
which involve several non--trivial aspects, such as the mixing between different
states, the Majorana nature of some particles and, if one aims to be rather 
precise, the higher order corrections which for the calculation of a single 
parameter need the knowledge of a large part of the remaining spectrum. One has
then to calculate in some accurate way, i.e. including higher order corrections,
the rates for the many possible decay modes and production processes at the 
various possible machines and eventually the implications for Dark Matter 
searches. \s

Fortunately, there are well motivated theoretical models where the soft 
SUSY--breaking parameters obey a number of universal boundary conditions at a high 
(unification) scale, leading to only a handful set of basic parameters. This is
the case for instance of the minimal Supergravity model (mSUGRA)
\cite{mSUGRA},  where the entire sparticle and Higgs spectrum is determined by
the values of  only five free parameters [a common gaugino mass $m_{1/2} =
M_i$, universal  scalar mass $m_0= m_{\tilde{f}}=m_{H_i}$ and trilinear
coupling $A_0=A_i$  at the GUT scale, the sign of the higgsino parameter $\mu$
and $\tb$, the  ratios of vevs of the two--Higgs doublets of the MSSM],  making
comprehensive scans of the parameter space  and detailed studies of the
spectrum feasible. However, there are also similarly constrained and highly
predictive models, such as anomaly  (AMSB) \cite{AMSB} and gauge  (GMSB)
\cite{GMSB} mediated SUSY-breaking model, string inspired models or  models
with right--handed neutrinos, to name a few, which can serve as benchmarks
\cite{benchmark} to be investigated. We then have to trade a complicated
situation where we have one general model  with many input parameters, with a
not less complicated situation where we  have many constrained models with a
small number of basic parameters. \s

In addition, in these unified models, the low--energy parameters are derived 
from the high--energy (GUT and/or possibly some intermediate scale) input
parameters through Renormalization Group Equations (RGE) and they should also
necessarily involve radiative electroweak symmetry breaking (EWSB), which sets
additional constraints.  The implementation of the RG evolution
and EWSB mechanism poses numerous non--trivial technical problems if they
have to be done accurately, i.e. including higher order effects. This
complication is to be added to the still present one stemming from the 
accurate calculation of the particle masses and couplings, decay and production 
rates, etc...\s

Therefore, to deal with the supersymmetric spectrum in all possible cases, one
needs very sophisticated programs to encode all the information and,
eventually, to pass it to Monte--Carlo event generators to simulate the
physical properties of the new particles.  These programs should have a high
degree of flexibility in the choice of the model and/or the input parameters
and an adequate level of approximation at different stages, for instance in the
incorporation of the RGEs, the handling of the EWSB  and the inclusion of the 
radiative corrections to (s)particle masses, which in many cases can be very
important. They should also be reliable, quite fast to allow for rapid
comprehensive scans of the parameter space and simple enough to be linked with 
other spectra programs or Monte--Carlo event generators.  There are several
public codes which deal with this topic and I will briefly discuss them
here\footnote{There are  also several private codes, dealing with one topic or
another, which I will not discuss here. However, I will mention a few still
private codes which will be  made public in a rather near future.}.

\subsection*{2. Codes for Spectra calculations}

There are four main public codes which make rather detailed calculations of the 
Supersymmetric particle spectrum in the pMSSM or in constrained  scenarii
(mSUGRA, etc..): 
\begin{itemize} 
\item[--] {\tt ISASUSY} \cite{ISAJET}, which is available since the early 90s 
and is implemented in the Monte--Carlo generator {\tt ISAJET}; 
it is the most widely used for simulations in the MSSM.\vspace*{-2mm}

\item[--] {\tt SuSpect} \cite{SUSPECT}, a new version has been released very 
recently but a preliminary version of the program exists since 
1998 and was described in Ref.~\cite{GDR}. 
\vspace*{-2mm}

\item[--] {\tt SOFTSUSY} \cite{SOFTSUSY}, a code written in C++ and has
been release a year ago. 
\vspace*{-2mm}

\item[--] {\tt SPHENO} \cite{SPHENO}, which is under development and will 
appear soon. 
\vspace*{-2mm}
\end{itemize}

The codes have different features in general, but they all incorporate the four
main ingredients or requirements for any complete calculation of the SUSY 
spectrum:\s

$i)$ RG evolution of parameters back and forth between the low energy scale, 
such as $M_Z$ and the electroweak symmetry breaking scale, and the high--energy
scale, such as the GUT scale or the messenger scale in GMSB models. This is the 
case for the SM gauge and Yukawa couplings and for the soft SUSY--breaking 
terms: scalar and gaugino masses, bilinear and trilinear couplings, the 
higgsino parameter $\mu$ and $\tb$, the ratios of vevs of the two--Higgs
doublets of the MSSM. This procedure has to be iterated several times to 
include SUSY threshold effects  or radiative corrections due SUSY particles.\s

$ii)$ The implementation of radiative EWSB and the calculation of the bilinear 
term $B$ and the absolute value of the higgsino parameter $|\mu|$ from the 
minimization of the full one--loop effective scalar potential [i.e. including 
all standard and SUSY particle loop contributions] at the EWSB scale which
provides two  additional constraints. The procedure has to be iterated until a
convergent  value for these parameters is obtained.\s 

$iii)$ Calculation of the pole masses of the Higgs bosons and all the
supersymmetric particles, including the possible mixing between the current 
states and the radiative corrections when they are important. An iteration, 
similar to the one for the RGEs, is also needed here to obtain the precise 
mass values.\s

$iv)$ The possibility of performing some checks of important theoretical 
features, such as the absence of tachyonic particles, non desired charge and 
color breaking (CCB) minima, a potential unbounded from below (UFB) and possibly
large fine--tuning in the EWSB conditions and also some experimental constraints
from negative searches of sparticles and Higgs bosons at colliders or from 
high precision measurements. \s

The general algorithm depicted in Figure 1, includes the various important 
steps:  the choice of SM inputs parameters at low energy [the gauge coupling
constant and the pole masses of the third generation fermions], the calculation
of the running couplings including radiative corrections in the modified 
Dimensional Reduction scheme $\overline{\rm DR}$ [which preserves SUSY] and 
their RG running back and forth between low and high scales, with the 
possibility of imposing the unification of the gauge couplings and the inclusion
of SUSY thresholds in some cases, the RG evolution of the soft--SUSY
breaking parameters from the high scale to the EWSB scale, the minimization of 
the one-loop effective potential and the determination of some important
parameters, and finally the calculation of the particle masses including the 
diagonalization of the mass matrices and the radiative corrections. \newpage

\begin{picture}(1000,470)(10,0)
\Line(390,450)(450,450)
\ArrowLine(450,-60)(450,450)
\Line(450,-60)(390,-60)

\Line(10,470)(390,470)
\Line(10,410)(390,410)
\Line(10,470)(10,410)
\Line(390,470)(390,410)

\Text(200,460)[]{Choice of low energy inputs:  $\alpha(M_Z), \sin^2\theta_W, 
\alpha_S(M_Z)$, $m_{t,b,\tau}^{\rm pole}$\, ; $\tan \beta (M_Z)$}

\Text(200,440)[]{Radiative corrections $\Rightarrow$ $g_{1,2,3}^{\rm 
\overline{DR}}(M_Z)$, $\lambda_\tau^{\rm \overline{DR}} (M_Z), \lambda_b^{\rm 
\overline{DR}}(M_Z), \lambda_t^{\rm \overline{DR}} (m_t)$}

\Text(200,420)[]{\it First iteration: no SUSY radiative corrections.} 

\ArrowLine(200,408)(200,392)

\Line(10,390)(390,390)
\Line(10,310)(390,310)
\Line(10,390)(10,310)
\Line(390,390)(390,310)

\Text(200,370)[]{Two--loop RGE for $g_{1,2,3}^{\overline{\rm DR}}$ and $\lambda_{\tau,b,t}^ 
{\overline{\rm DR}}$ with choice: $\begin{array}{l}  g_1=g_2 \cdot \sqrt{3/5} \\ 
 M_{\rm GUT} \sim 2 \cdot 10^{16}~{\rm  GeV} \end{array}$}

\Text(200,340)[]{Include all SUSY thresholds via step functions in $\beta$ 
functions.}

\Text(200,320)[]{\it First iteration: unique threshold guessed.} 

\ArrowLine(200,308)(200,295)

\Text(200,285)[]{Choice of SUSY-breaking model (mSUGRA, GMSB, AMSB, or pMSSM).}
 
\Text(200,265)[]{Fix your high--energy inputs (mSUGRA: $m_0, m_{1/2}, A_0$, 
sign($\mu)$, etc...).} 

\ArrowLine(200,253)(200,242)

\Line(10,240)(390,240)
\Line(10,180)(390,180)
\Line(10,240)(10,180)
\Line(390,240)(390,180)

\Text(200,220)[]{Run down with RGE to: $\begin{array}{l} - M_Z(m_t)~{\rm for}~
g_{1,2,3}~{\rm and}~\lambda_{\tau,b}(\lambda_t) \\
- M_{\rm EWSB}~{\rm for}~\tilde{m}_i, M_i, A_i, \mu, B \end{array}$} 
\Text(200,195)[]{\it First iteration: guess for $M_{\rm EWSB} = M_Z$.}

\ArrowLine(200,178)(200,162)

\Line(10,160)(390,160)
\Line(10,100)(390,100)
\Line(10,160)(10,100)
\Line(390,160)(390,100)

\Text(200,150)[]{$\mu^2, \mu B = F_{\rm non-linear}(m_{H_1}, m_{H_2}, 
\tan\beta, V_{\rm loop} )$}

\Text(200,130)[]{$V_{\rm loop} \equiv $ Effective potential at 1--loop with 
all masses.}

\Text(200,110)[]{{\it First iteration: $V_{loop}$ not included} }

\Line(390,130)(420,130)
\ArrowLine(420,30)(420,130)
\Line(420,30)(390,30)

\ArrowLine(200,98)(200,90)

\Text(200,85)[]{Check of consistent EWSB ($\mu$ convergence, no tachyons, 
simple CCB/UFB, etc...) } 

\ArrowLine(200,75)(200,62)

\Line(10,60)(390,60)
\Line(10,0)(390,0)
\Line(10,60)(10,0)
\Line(390,60)(390,0)

\Text(200,50)[]{Diagonalization of mass matrices and calculation of masses /  
couplings}

\Text(200,30)[]{Radiative corrections to the physical Higgs, sfermions, gaugino
masses.}

\Text(200,10)[]{\it First iteration: no radiative corrections.} 

\ArrowLine(200,-2)(200,-17)

\Line(10,-20)(390,-20)
\Line(10,-85)(390,-85)
\Line(10,-20)(10,-85)
\Line(390,-20)(390,-85)

\Text(200,-30)[]{Check of a reasonable spectrum:} 

\Text(200,-45)[]{-- no tachyonic masses (from RGE, EWSB or mix), 
good LSP, etc..\ \hfill } 

\Text(200,-60)[]{-- not too much fine-tuning and sophisticated CCB/UFB 
conditions,\hfill }

\Text(200,-75)[]{-- agreement with experiment: $\Delta\rho$, $(g-2)$, 
$b \to s\gamma$. \ \ \ \ \ \ \ \ \ \ \ \ \ \ \ \ \ \ \ \ \ \hfill}
\end{picture}

\vspace*{3.5cm}

\nn {\it Figure 1: Iterative algorithm for the calculation of the SUSY particle
spectrum  from the choice of inputs to the check of the spectrum. 
The small iteration on $\mu$ is performed until $\mu_{i}-\mu_{i-1} \leq \epsilon$
while the long RG/RC iteration needs to be performed at least 3 to 4 
times. In the first iteration, no radiative corrections are included and the 
SUSY thresholds as well as the EWSB and GUT scales are guessed.}\bigskip  

For the various aspects of the calculation, the previous four codes have 
different features in general [which is very useful for performing 
cross--checks]: some are written in Fortran and some in C++,  some are
interfaced with  event generators or other programs,  they have different
options for models  and input parameters (flexibility) and they use different
approximations in  the calculation (for instance in the inclusion of the higher
order radiative  corrections, the RGE running, the EWSB mechanism, etc..). In
Table 1, the  various features of the four programs are summarized. As can be
seen, they all  deal with the most studied theoretical models [mS, AM and GM
stand,  respectively, for mSUGRA, AMSB and GMSB], use sometimes different 
approximations [although not in very important sectors such as the gauge and 
Yukawa couplings] and many calculate additional items and/or are interfaced 
with other programs. 

\begin{table}[h!]
\renewcommand{\arraystretch}{1.3}
\vspace*{-.3cm}
\begin{center}
\begin{tabular}{|c|c|c|c|c|}
\hline 
Item/Code  & {\tt ISASUSY} & {\tt SuSpect} & {\tt SOFTSUSY} & {\tt SPHENO} \\ 
\hline
Language & Fortran & Fortran & C++ & Fortran \\ \hline \hline 
Models  &mS,AM,GM& mS,AM,GM & mS,AM,GM& mS,AM,GM \\ 
       &pMSSM(25)& pMSSM(22)& -- & -- \\ 
       & $\tilde{\nu}_R$, strings & -- & -- & string sc. \\ \hline
RGEs  & 2--loop $g_i,\lambda_i$ & 2--loop $g_i,\lambda_i$ &
2--loop $g_i,\lambda_i$ & 2--loop $g_i,\lambda_i$ \\
    & 2 loop soft & 1--loop soft & 1--loop soft & 2--loop soft 
\\ \hline
EWSB  & $\sqrt{m_{{\tilde t}_1} m_{{\tilde t}_2}}$ & flexible & $\sqrt{m_{{\tilde t}_1} m_{{\tilde t}_2}}$ & 
$\sqrt{m_{{\tilde t}_L} m_{{\tilde t}_R}}$ \\ 
$V_{\rm loop}$/tad. & $t,b,\tilde{t},\tilde{b}$ & 1--loop & 1--loop & 1--loop 
\\ \hline
Thresholds  & Steps & Steps & in RC & in RC \\ \hline
\ \ \  \ \ \ \ SM & leading & lead/full & lead/full & full \\
RC \ SUSY &   approx.& $\sim$ PBMZ  & $\sim$ PBMZ  & full \\ 
\ \ \ \ \ \ \ Higgs & 1Loop EP & SUBH/FHF/HHH & FHF & BDSZ \\ \hline 
Checks       & -- & CCB,UFB,FT    & FineTuning        & CCB,UFB \\
       & -- & EW,$a_\mu$,$bs\gamma$ & -- & EW,$a_\mu, b s\gamma$  \\ \hline \hline 
Decays & Yes & HDECAY/SDECAY$^*$& -- & Yes \\ \hline 
Production & $pp$ and $e^+e^-$ & $ee$ SUSYGEN, $pp^*$ & -- & $e^+e^-$ \\ \hline
DM calc. & -- & $\mu$Megas/DarkSUSY  & $\mu$Megas & --  \\ \hline
\end{tabular}
\end{center}
\vspace*{-.5cm}
\end{table}
\nn {\it Table 1: The various important items implemented in the four RGE 
codes for the calculation of the (s)particle spectra. The $^*$ 
means that the item is under implementation.}\bigskip

One important ingredient in these calculations is the radiative corrections to
(s)particle masses. These corrections, in particular those stemming from QCD 
and third generation (s)fermions because of the strong couplings, can be 
large. In the SUSY sector, they can alter the experimental search strategies at 
colliders since large corrections may change constraints on the MSSM parameter 
space and may allow or not some new decay modes. Moreover, the mass 
difference between the lightest neutralino [which is in general the LSP]
and the other sparticles plays a major role, since it gives the amount of missing 
energy [which is the typical signature of SUSY processes at colliders] and 
enters Dark Matter relic density calculations [co--annihilation]. In most of 
the codes above, these radiative corrections are implemented {\it \`a la} 
Pierce, Bagger, Matchev and Zhang (PBMZ) \cite{PBMZ}. \s

The radiative corrections are particularly important, as is well known, in 
the Higgs sector where one-- and two--loop contributions of third generation
(s)fermions can shift the upper limit on the mass of the lightest Higgs
boson from $M_Z$ by up to 40 GeV, dramatically changing the reach of the 
LEP2 collider for instance. In fact, this is the most delicate quantity 
to calculate. RGE codes such as {\tt ISASUSY} and {\tt SuSpect} have 
their own approximate calculations, but they are also linked with routines 
which do a more sophisticated job. The main available routines for the Higgs 
sector are: 

\begin{itemize} 
\item[--] {\tt Subhpole} (SUBH)  \cite{SUBH}: which calculates the leading 
radiative corrections in the effective potential approach with a two--loop
RG improvement. It includes the leading $\lambda_t^2$ corrections as
well as the leading part of the SUSY--QCD corrections. 
\vspace*{-2mm}

\item[--] {\tt HMSUSY} (HHH) \cite{HHH}: calculates the one--loop corrections
in the effective potential approach and includes the leading two--loop standard
QCD and EW corrections. 
\vspace*{-2mm}

\item[--] {\tt FeynHiggsFast} (FFH) \cite{FeynHiggsFast}: calculates the 
corrections in the Feynman diagrammatic approach with the one and two--loop 
QCD corrections at zero momentum transfer [the version {\tt FeynHiggs} 
\cite{FHF} has the full one--loop corrections and is slower].  
\vspace*{-2mm}

\item[--] {\tt BDSZ} \cite{BDSZ} gives the leading one--loop corrections from
the third generation (s)fermion sector as well as the full $\alpha_s\lambda_t^2,
\lambda_t^4$ and $\alpha_s \lambda_b^2$ corrections at zero--momentum 
transfer.\vspace*{-2mm}
\end{itemize}

Detailed comparisons of these codes have been performed. The main conclusion
is  that despite of the different ways the various items discussed above are 
implemented, they in general agree at the percent level in large parts of the 
MSSM parameter space. Several more important differences occur however in some 
areas of the parameter space, in particular in the high $\tb$ and/or focus
point regions with large $m_0$ values, where the Yukawa couplings of top and
bottom quarks  play an important role\footnote{For a comparison of the four
codes, see the talk of Sabine Kraml at  this conference \cite{comp}. Note that
the comparison with the program {\tt  SuSpect} made there was with an earlier
version which had only a very  approximate determination of the Higgs boson
masses. The new version, since it  is linked to several Higgs routines, gives a
much better determination of these parameters.}. \s

\setcounter{figure}{1}
\begin{figure}[htbp]
\hspace*{2.5cm}{\large $m_0$}\\[-1.5cm]
\begin{center}
\hspace*{-.2cm} \epsfig{figure=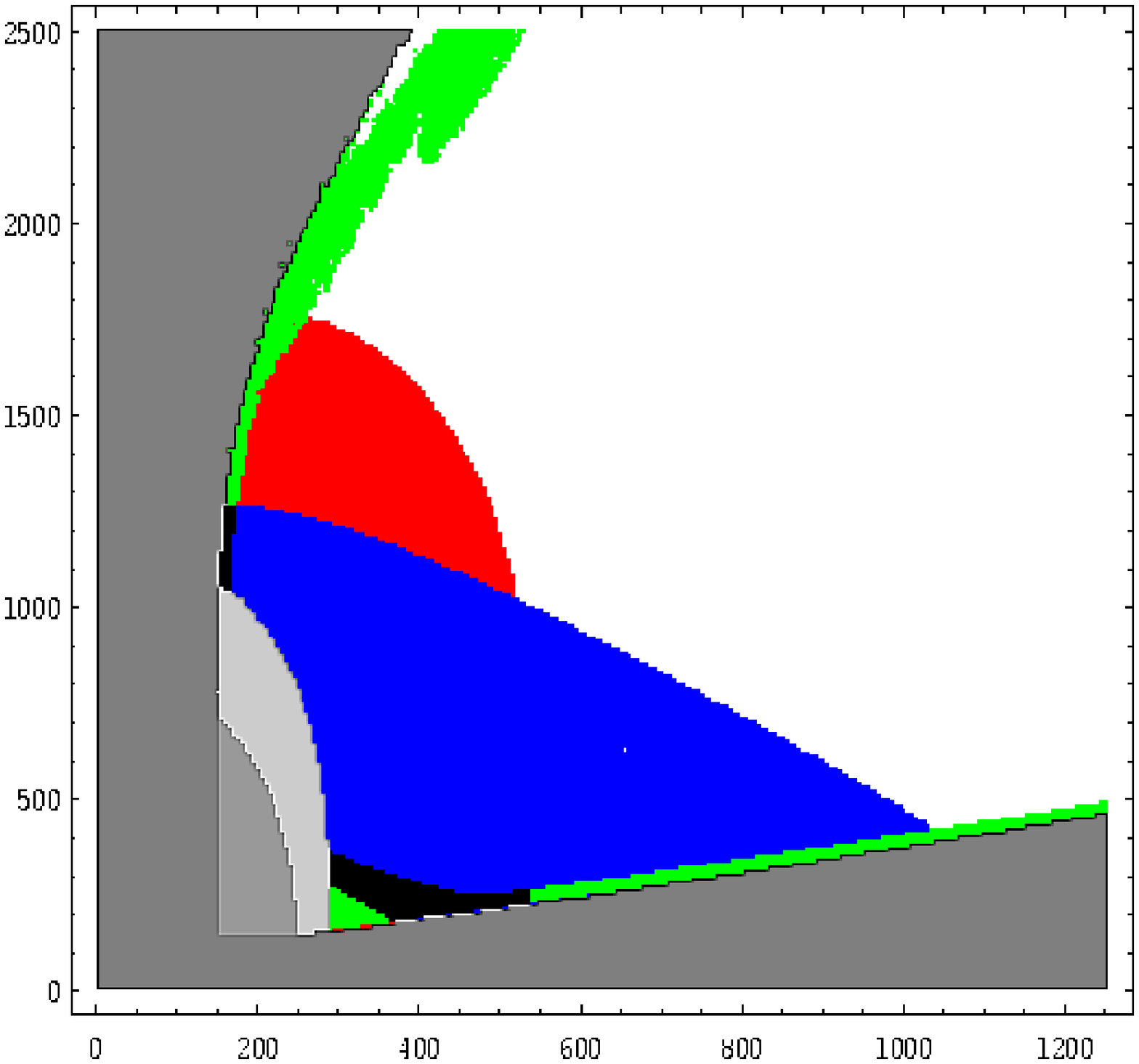,width=8.5cm,angle=-90}\\
\end{center}
\vspace*{-1cm}
\hspace*{12.3cm} {\large $m_{1/2}$}
\vspace*{1mm}
\caption[]{\it Constraints on the $(m_{1/2}, m_0)$ mSUGRA plane for 
$\tan \beta=40$, $A_0=0$ and sign$(\mu)>0$. The gray areas 
are those excluded by the requirement of EWSB and limits on SUSY 
particle masses (darker gray), BR($b \to s \gamma)$ (medium Rey) and 
$M_h > 113$ GeV (light and dark Frey). The colors are for the possible 
``evidence" for the LEP2 Higgs boson (red), the $(g_\mu-2)$ excess (blue) and 
the LSP being the Dark Matter (green).}
\end{figure}

Once the spectrum is calculated, one has the possibility of linking the 
previous programs with other routines which determine some properties of the
SUSY particles, imposing theoretical and experimental constraints and  making
scans on the parameter space to constrain the various models or to  delineate
regions of the parameter space where SUSY signals can be expected in colliders
or DM searches. This is exemplified in Fig.~2 from  Ref.~\cite{DDK}, where in
the mSUGRA model a scan in the $(m_0, m_{1/2})$  plane has been performed [for
given values of the other parameters $\tb, A_0$ and sign($\mu)$] including
theoretical constraints [proper EWSB, no CCB/UFB/tachyons, $\chi_1^0$ LSP] as
well as experimental constraints [bounds on sparticle and Higgs masses and
precision measurement from LEP and  the Tevatron, the decay $b \to s\gamma$]
and possibly some additional  requirements [such as the  $2\sigma$ evidence for
a 115 GeV SM Higgs at LEP, a 3 $\sigma$ contribution to the $(g-2)_\mu$ and a
$\chi_1^0$ LSP being a solution  for the DM problem i.e. with the relevant
relic density].

\subsection*{3. Codes for Production, Decay and Dark Matter calculations}

\subsubsection*{3.1 NLO Higgs and SUSY particle production calculations}

The incorporation of next--to--leading order (NLO) corrections is very 
important for Higgs boson and SUSY particle production at high--energy 
colliders. In particular, the QCD corrections at hadron machines LHC  or
Tevatron, can be rather significant with K--factors  as large as two
\cite{michael}. In  turn, radiative corrections in $\ee$ collisions processes
are in general much smaller but they can be measurable \cite{walter}. Here is
a  non--exhaustive list of available public codes for Higgs and sparticle 
production at NLO or including higher order effects:\s

\nn $\bullet$ NLO MSSM Higgs boson production at hadron colliders 
\cite{michael}:

-- {\tt HIGLU}: for the loop induced Higgs production $pp \to gg \to h,H,A$ 
(NLO).

-- {\tt VV2H/V2HV}: for production with gauge bosons $qq \to h,H + qq $ 
and $W,Z$ (NLO).

-- {\tt HQQ}: for radiation off top quark $pp \to q\bar{q},gg 
\to h,H,A+Q\bar{Q}$ (LO, NLO to come). 

-- {\tt HPAIR}: for Higgs pair production $pp \to q\bar{q},gg \to 
hh,HH,hA,HA,AA$ (partly NLO).\s

\nn $\bullet$ NLO SUSY particle production at hadron colliders \cite{michael}:

-- {\tt PROSPINO}: for squark and gluino production $pp \to \tilde{q} 
\tilde{q}^*, \tilde{q}\tilde{q}, \tilde{q}\tilde{g}, \tilde{g}
\tilde{g}$ at NLO. 

-- The pair and associated production of gauginos at NLO is under preparation. 
\s

\nn $\bullet$ Production of Higgs and SUSY particles at $\ee$ colliders: 

-- {\tt SUSYGEN} \cite{SUSYGEN} for Higgs and sparticle production, also a MC 
generator (see later).

-- {\tt HZHA} \cite{HZHA}: the most used Monte Carlo generator for Higgs 
production at LEP2. 

-- Many four or six fermion production processes at $\ee$ colliders....

\subsubsection*{3.2 Decays of Higgs and SUSY particles}

The decays of SUSY and Higgs particles can be rather complicated and it is 
important to determine them with a good accuracy. There can be a large number
of decay modes for some particles: simple two--body decays in which it is
important sometimes to include higher order corrections [as is the case  for
Higgs bosons and strongly interacting sparticles], and rather complicated
many--body decay modes such as the three or four body decays of charginos, 
neutralinos and top squarks or important loop--induced decay modes. There are
several  available codes,  doing this job with a different level of
sophistication: \s

-- {\tt ISASUSY} \cite{ISAJET}: only tree--level two--body Higgs and 
SUSY decays (3 body for gauginos).  

-- {\tt HDECAY} \cite{HDECAY}: SM and MSSM Higgs decays with higher order 
effects. 

-- {\tt SDECAY} \cite{SDECAY}: sparticle decays including higher order 
effects (RC and multi--body). 

-- {\tt SPHENO} \cite{SPHENO}: discussed above and has 2 and 3--body SUSY 
particle decays. \s

\nn Some decay routines are also included in the Monte--Carlo event generators 
{\tt SUSYGEN} \cite{SUSYGEN}, {\tt HZHA} \cite{HZHA}, {\tt PYTHIA} 
\cite{pythia} and {\tt HERWIG}  \cite{herwig} with possible links to the
programs mentioned above. Some  development in this subject is expected in the
near future. 

\subsubsection*{3.3 Dark Matter Codes}

Several experiments for cold Dark Matter searches are in progress or are planed
for a near future. The MSSM has a very good candidate, the LSP neutralino 
$\chi_1^0$, which is electrically neutral, weakly interacting, massive, 
absolutely stable and which can have the proper cosmological relic density 
\cite{LSP}. An intensive phenomenological activity is happening in this field 
and numerous analyses in the (un)constrained MSSM are performed for: \s

-- the relic density of the LSP: $\sigma(\chi_1^0 \chi_1^0+\chi_1^0  \tilde{P}
+  \tilde{P} \tilde{P} \to $ anything),
 
-- the rate for the direct detection of the LSPs: $\chi_1^0 N \to \chi_1^0 N$,

-- the rate for indirect detection: $\chi_1^0 \chi_1^0 \to \gamma \gamma, 
\gamma Z$ and $\bar{p}, e^+, \nu+X$. \s

For the calculation of the relic density, one needs the sparticle and Higgs 
spectra and couplings, the annihilation and co--annihilation cross sections, 
pair production thresholds, effects of resonances, etc... For the detection, 
one also needs the modelling of the halo, the hadronization, the nuclear matrix 
elements, the particle flux, interaction and propagation, etc... There are two 
main multi--purpose codes which include all these\footnote{See the talk of 
Emmanuel Nezri in the parallel sessions \cite{Nezri}.}:  \s

\begin{itemize}
\vspace*{-3mm} 
\item[--] {\tt DarkSUSY} \cite{darksusy}: which has its own pMSSM spectra 
calculation but can be be linked to {\tt SuSpect}.  The hadronization and SM 
particle decays are taken from {\tt PYTHIA}. The program is widely used,
in particular for the indirect detection rates. 
\vspace*{-2mm}

\item[--] {\tt NeutDriver} \cite{neutdriver}: which has only implemented the 
unconstrained  MSSM with 69 parameters and the obtained spectrum seems to be 
problematic. To my knowledge, there was no recent upgrade of the program.
\vspace*{-2mm}
\end{itemize}

There is also a new code {\tt micrOMEGAs} \cite{micromegas}, which  calculates
the relic density in the constrained or pMSSM. The (co--)annihilation
calculation with all channels included is based on {\tt CompHEP} \cite{Comphep}
and includes links to {\tt ISASUSY, SuSpect, HDECAY} for the calculation of the
spectra. A number of private codes for one or all of these items also
exist\footnote{In  particular the codes by A. Arnowitt et al. and K. Olive et
al. on which some  recent MSSM Dark Matter analyses discussed at this
conference are based upon.}.

\subsection*{4. Automatic Matrix Element Generators}

Processes in which there are many particles in the final state are very 
important for $pp$ and $\ee$ physics. For instance, the $pp$ or $\ee \to H t 
\bar{t} \to$ process, which allows to measure the top Yukawa coupling, leads to
8 or 10 final fermions depending on whether the Higgs decays into  $b\bar{b}$ 
or $WW$ pairs. The full processes have very large matrix elements and in many
cases, they  need to be calculated automatically and interfaced to MC event 
generators for a full simulation. There are several codes available on the
market for SUSY and SM processes [the latter being needed for the calculation
of the backgrounds]:

\begin{itemize}
\vspace*{-3mm} 
\item[--] {\tt CompHEP} \cite{Comphep}: is one of the major codes for matrix
elements calculations. It uses trace techniques for the algebra, Vegas for
phase-space integration and calculates its own SUSY Feynman rules and spectra. 
It has an easy interface with MC generators {\`a la} ``Les Houches accord" 
\cite{accord} and the program is developing quite rapidly.
\vspace*{-3mm} 

\item[--] {\tt GRACE--SUSY} \cite{grace}: it has for the moment only  $\ee$ 
production processes, only selected processes and needs model files for the 
others. It uses Form and Reduce for trace calculations. There was no recent 
major SUSY development but it has been recently used to calculate the 
${\cal O}(\alpha)$ corrections to $\ee \to \nu \bar{\nu}H$ in the SM 
\cite{nnH}.
\vspace*{-3mm}

\item[--] {\tt FeynCalc} \cite{FeynCalc}: together with {\tt FeynArts} for
the drawing of the Feynman diagrams is mostly used for loop calculations
in the SM and the MSSM.  
\vspace*{-3mm}

\item[--] {\tt AMEGIC++} \cite{amegic++}: is a C++ program for multi--particle
production (no calculation of loops yet). It is now implementing SUSY
processes in $\ee$ collisions. 
\vspace*{-2mm}
\end{itemize}
There exist other codes multi--particle production codes, such as {\tt 
O'MEGA/WHIZARD} \cite{omega} and {\tt MadGraph} \cite{madgraph} for instance,
but they do not include SUSY processes yet. 

\subsection*{5. Monte--Carlo event generators} These are big mastodons which
are not specific to SUSY and which perform all analyses from the production of
the (new) particles to the hadron decays and simulate the signals and the
various backgrounds. They in general include five main phases in the simulation
process: 1) the hard production processes,  2) the parton showering, 3) the
heavy particle decays, 4) the hadronization process and 5) the hadron decays.
In the following, I will briefly discuss only the parts on the production (1) 
and decays (3) of SUSY/Higgs particles; the rest of the  simulation is as in the 
Standard Model.  There are three plus one multipurpose Monte--Carlo event 
generators\footnote{ Thanks to Peter Richardson for his advice in this
section.}:

\begin{itemize}
\vspace*{-2mm}
\item[$\bullet$] {\tt ISAJET} \cite{ISAJET}: this is the oldest and most used of
all generators dealing with SUSY processes. It has  all SUSY production
channels (including  some $R_p \hspace*{-4mm}/$ \ violating processes) and is
linked with {\tt  ISASUSY} which is built in for the spectrum and decay
branching ratios  calculation. The known drawback is that it has not a 
very satisfactory description of the standard processes (steps  2, 4 and 5).  
\vspace*{-2mm}

\item[$\bullet$] {\tt (S)PYTHIA} \cite{pythia}: it is known to give one of the 
best description of SM physics. It has its own calculation of the 2--body
decay  rates of SUSY and Higgs particles and has implemented a  wide range of 
production processes (including $R_p \hspace*{-4mm}/$ processes). But it  has
not a very precise determination of the Higgs and SUSY particle spectrum  since
it is based on analytical formulae which are rough approximations.   
\vspace*{-2mm}

\item[$\bullet$] {\tt HERWIG} \cite{herwig}: which is very good to describe the
SM and QCD aspects but the SUSY aspect of the program is developing very 
rapidly. It has many production channels, a good treatment of $R_p 
\hspace*{-4mm}/$\ \, ,includes the spin correlations in most processes and 
allows for the polarization of the initial beams in $\ee$ collisions. There  
is built--in code for spectra or decay calculation but has an interface with 
{\tt ISAJET}, and in a near future {\tt SuSpect} and {\tt HDECAY/SDECAY}.
\vspace*{-2mm}

\item[$\bullet$] {\tt SUSYGEN} \cite{SUSYGEN}: which is specialized in $\ee$ 
collisions (in particular it was used to describe LEP physics) but includes 
now some processes in $pp$ and $ep$ collisions. The spectrum calculation is 
performed by {\tt SuSpect} and  for the decay widths and branching ratios 
it is linked to {\tt HDECAY} for the Higgs sector while it has its own 
calculation for the SUSY sector. It has full spin correlations and includes 
most $R_p \hspace*{-4mm}/$\ \ processes. It is interfaced with {\tt PYTHIA} for
parton shower and hadronization since it cannot simulate the SM backgrounds. 
\vspace*{-2mm}
\end{itemize}

\subsection*{6. Conclusions}

In the recent years, more and more programs for phenomenological and 
experimental analyses became available and the tendency to make them public is
growing rather fast. This is very useful for the reliability of these tools 
since many checks and comparisons can be then performed, thus  minimizing the 
number of errors, bugs and inconsistencies. It also generates a healthy 
competition between the various codes which are more often upgraded to take
into account new developments. The programs are  becoming more and more
sophisticated but at the same time, efforts are devoted to make them more
clear,  user friendly and with the adequate documentation. \s

Due to the complexity of the subject, most programs deal with only one or a few
aspects of the theoretical, phenomenological or experimental facets of SUSY. 
This calls for complementarity between the various programs for spectra
determination, higher order corrections, matrix elements calculations,  Dark
Matter analyses, Monte--Carlo event generators, etc... A large ``communication"
effort is  made: many workshops devoted to tools are organized
[GDR--Supersym\'etrie, and  many discussions to complement and interface the
various programs are taking  place [see for instance the ``accord"
\cite{accord} obtained during the Les  Houches Workshop]. This leads to more
interplay between theory and experiment  which is very useful for the field. \s

To summarize, there is a very rapid development of the field. In addition to
the fact that many new codes have appeared and many developments in the major
spectra codes and Monte--Carlo generators have occurred in the recent years, 
the trend is to make them  much faster, more efficient, user--friendly  and as 
complete as possible. Many programs are moving to C++, although there is still 
a lack of a complete consensus whether it is mandatory [probably a reflection 
of the conflict between generations..]. Therefore, we will be certainly ready 
to analyse the data for the next round of experiment. The hope is that SUSY is 
also ready to be discovered! \bigskip

\nn {\bf Acknowledgements:} I thank the organizers of this conference, in
particular Peter Zerwas, for their invitation and for the stimulating atmosphere.
Thanks also go to Jean-Lo\"{\i}c Kneur, Sabine Kraml and Peter Richardson for
providing me with some material.

\end{document}